\documentclass[twocolumn,10pt]{article}
  
\usepackage{graphicx}
\usepackage{dcolumn}
\usepackage{bm}
\usepackage{bbm}
\usepackage{amsfonts}

\def\>{\rangle}
\def\<{\langle}
\def\ket#1{|#1\>}
\def\bra#1{\<#1|}

\def\ii{{\rm i}}
\def\dd{{\rm d}}

\begin{document}

\vbox{
\onecolumn

\noindent
{{\LARGE\bf Quantum chaos and the double-slit experiment}}
\bigskip
\bigskip

\noindent
{\bf Giulio Casati$^{1,2,3,5}$ and Toma\v z Prosen$^{4,5}$}
\bigskip

\noindent
{\em\small
$^1$ Center for Nonlinear and Complex Systems, Universita' degli Studi dell'Insubria, 22100 Como, Italy\\
$^2$ Istituto Nazionale per la Fisica della Materia, unita' di Como, 22100 Como, Italy,\\
$^3$ Istituto Nazionale di Fisica Nucleare, sezione di Milano, 20133 Milano, Italy,\\
$^4$ Department of Physics, Faculty of mathematics and physics, University of Ljubljana, 1000 Ljubljana, Slovenia,\\
$^5$ Department of Physics, National University of Singapore, Singapore
}

\bigskip

{\bf 
We report on the numerical simulation of the double-slit experiment,
where the initial wave-packet is bounded inside a billiard domain with 
perfectly reflecting walls. If the 
shape of the billiard is such that the classical ray dynamics is regular,
we obtain interference fringes whose visibility can be controlled
by changing the parameters of the initial state.
However, if we modify the shape of the billiard
thus rendering classical (ray) dynamics fully chaotic, the
 interference fringes disappear and the intensity on the screen becomes the
(classical) sum of  intensities for the two corresponding one-slit
experiments.
Thus we show a clear and fundamental example in which transition to chaotic motion in a deterministic classical system, in absence of any external noise, leads to a profound modification in the quantum behaviour.
}

}
\twocolumn
\section*{Introduction}

As it is now widely recognized, classical dynamical chaos has been one of the 
major scientific breakthroughs of the past century. On the other hand, the 
manifestations of chaotic motion in quantum mechanics, though widely 
studied \cite{haake,stockmann}, 
remain somehow not so clearly understood, both from the mathematical as well 
as from the physical point of view.

The difficulty in understanding chaotic motion in terms of quantum mechanics 
is rooted in two basic properties of quantum dynamics:
\begin{enumerate}
\item
The energy spectrum of bounded, finite number of particles, 
conservative quantum systems 
is discrete. This means that the quantum motion is ultimately
quasi-periodic, i.e. any temporal behaviour is a discrete superposition 
of finitely or countably many Fourier components with discrete frequencies.
In the ergodic theory of classical dynamical systems, 
such a quasi-periodic dynamics corresponds to the limiting case of 
integrable or ordered motion 
while chaotic motion requires continuous Fourier spectrum \cite{sinai}.
\item
Quantum motion is dynamically stable, i.e. initial errors propagate only 
linearly with time \cite{casati}. Linear instability is a typical feature of classical 
integrable systems and this contrasts the exponential instability which 
characterizes classical chaotic systems.
\end{enumerate}
Therefore it appears that quantum motion always exhibits the characteristic 
features of classical integrable, regular motion which is just the opposite of 
dynamical chaos.
However, it has been shown that this apparently paradoxical situation can be 
resolved with the introduction of different time scales inside which the 
typical features of classical chaos are present in the quantum motion also. 
Since these time scales diverge as Planck constant $\hbar$ goes to zero, no 
contradiction arises with the correspondence principle \cite{boris}.

Still the state of affairs remains unsatisfactory. For example one should 
build a statistical theory for systems with discrete spectrum and linear 
instability. In this connection the question whether, in order to have the 
quantum to classical transition, external noise (or coupling to external
macroscopic number of degrees of freedom) is necessary or not, remains 
unclear. Indeed it is generally accepted that external noise may induce the 
non-unitary evolution leading to the decay of non-diagonal matrix elements
of the density matrix in the eigenbasis of the physical observables, 
thus restoring the classical behaviour.
On the other hand it has also been surmised that external noise, being 
sufficient, is not necessary. A new type of decoherence -- 
the {\it dynamical decoherence}-- has been proposed \cite{boris}, 
without any noise and only due to the 
intrinsic chaotic evolution of a pure quantum state. 
The simplest manifestations of dynamical decoherence are the fluctuations 
in the quantum steady state which, in the quasi-classical region, 
is a superposition of very many eigenfunctions. 
In case of a quantum chaotic - ergodic steady state - all 
eigenfunctions essentially contribute to the fluctuations and their 
contribution is statistically independent\cite{boris}. This fact suggests the 
complete quantum decoherence in the final steady state for any initial
 state even though the steady state is formally a pure quantum state. 
Yet this argument is not completely convincing and a more clear evidence 
is required.
In this letter we discuss this question by considering one of the basic 
experiments on which rests quantum mechanics, namely a phenomenon which, 
in the words of Richard Feynmann \cite{feynmann}, 
"... is impossible {\it absolutely} impossible, 
to explain in any classical way, and which has in it the heart of quantum 
mechanics. In reality, it contains the {\it only} mystery." : the double slit 
experiment.

\section*{Experiment}

\begin{figure}
\centerline{\includegraphics[width=3.3in]{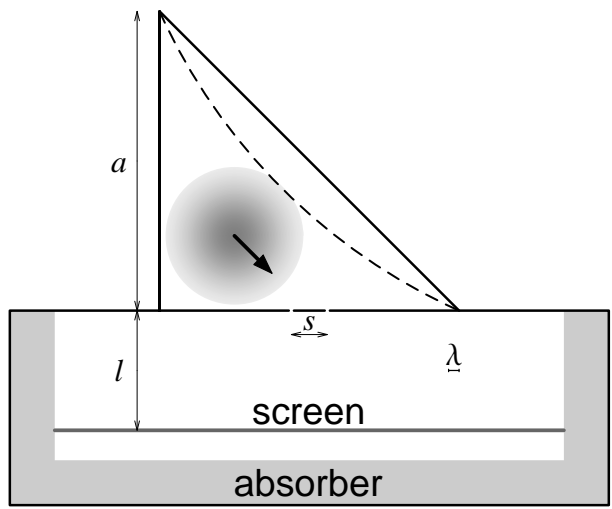}}
\caption{The geometry of the numerical double-slit experiment. All scales 
are in proper proportions. The two slits are placed at a distance $s$ on the lower side of the billiard
}
\label{fig:1}
\end{figure}

We have performed the following numerical, double-slit experiment. The time dependent Schr\" odinger equation 
$\ii\hbar \frac{\partial}{\partial t} 
\Psi(x,y,t) ={\hat H} \Psi(x,y,t)$, 
with ${\hat H} = {\hat p}^2/(2m)$, 
has been solved numerically (see {\em Method} section)
for a quantum particle which moves freely inside the 
two-dimensional domain as indicated in fig.~\ref{fig:1} (full line). 
Note that the 
domain is composed of two regions which are connected only through two narrow
slits. We refer to the upper bounded region as to the {\em billiard domain}, 
and to the lower one as the {\em radiating region}. The scaled units 
have been
used in which Planck's constant $\hbar=1$, mass $m=1$, and the base of the
triangular billiard has length $a=1$.
The initial state $\Psi(t=0)$ is a Gaussian wave packet (coherent state)
centered at a distance $a/4$ from the lower-left 
corner of the billiard (in both Cartesian directions)
and with velocity $\vec{v}$ pointing to the middle 
between the slits. The screen is at a distance $l=0.4$ from the base of the
triangle. 
The magnitude of velocity $v$ (in our units equal to the wave-number $k=v$)
sets the de Broglie wavelength $\lambda=2\pi/k$. In our experiment we have
chosen $k=180$ corresponding to approximately $1600$th excited
states of the closed quantum billiard.
The slits distance has been
set to $s=0.1 \approx 3 \lambda$ and the width of the slits is 
$d=\lambda/4$. The wave-packet is also characterized by the
position uncertainty $\sigma_x=\sigma_y=0.24$. This was chosen 
as large as possible in the present geometry
in order to have a small uncertainty in momentum $\sigma_k = 1/(2\sigma_x)$.

The lower, radiating region, should in principle be infinite. 
Thus, in order to efficiently damp waves at finite
boundaries, we have introduced an absorbing layer around the radiating region. 
More precisely, in the region referred to as absorber, we 
have added a negative imaginary potential to the Hamiltonian 
$H \to H - \ii V(x,y), V\ge 0$, 
which, according to the time dependent Schr\" odinger equation, 
ensures exponential damping in time. 
In order to minimize any possible reflections from the
border of the absorber, we have chosen $V$ to be smooth, starting from zero
and then growing quadratically inside the absorber. No significant
reflection from the absorber was detected and this ensures that the
results of our experiment are the same as would be for an infinite 
radiating region.

\begin{figure}
\centerline{\includegraphics[width=3.3in]{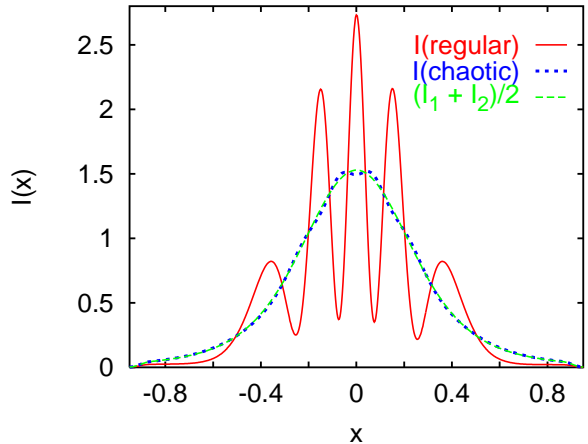}}
\caption{
The total intensity after the double-slit experiment as a function of the 
position on the
screen. $I(x)$ is obtained as the perpendicular component of
the probability current, integrated in time. The red full curve
indicates the case of regular billiard, while the blue dotted curve
indicates the case of chaotic one.
The green dashed curve indicates the averaged intensity over two 1-slit
experiments, with either the regular or chaotic billiard (with
results being practically the same, see fig.~\ref{fig:1slit}).
}
\label{fig:main}
\end{figure}

\begin{figure}
\centerline{\includegraphics[width=3.3in]{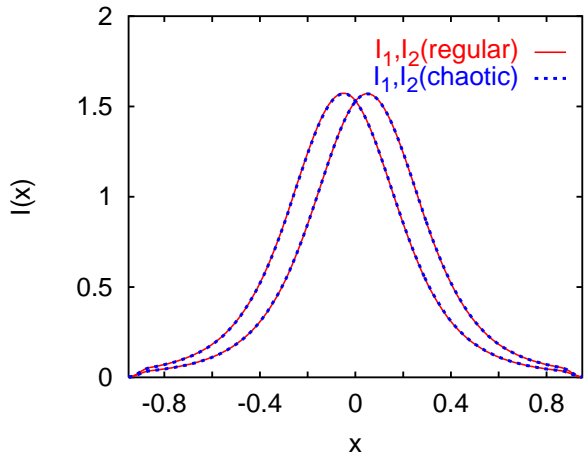}}
\caption{
The two pairs of curves represent the intensities on the screen for the two
1-slit experiments (with either one of the two slits closed).
The red full curves indicate the case of the regular billiard while
the blue dotted ones indicate the case of chaotic billiard.
}
\label{fig:1slit}
\end{figure}

While the wave-function evolves with time, a small probability current
leaks from the billiard and radiates through the slits. The radiating
probability is recorded on a horizontal line $y = -l$ referred to
as the screen. The experiment stops when the probability that the
particle remains in the billiard region becomes vanishingly small.
We define the intensity at the position $x$ on the screen as the 
perpendicular component of the probability current, integrated in time
\begin{equation}
I(x) = \int_0^\infty \!\!\dd t\,
{\rm Im}\, 
\Psi^*(x,y,t)\frac{\partial}{\partial y}\Psi(x,y,t)\vert_{y=-l}.
\end{equation}
By conservation of probability the intensity is normalized, 
$\int_{-\infty}^\infty \dd x I(x) = 1$, and is positive $I(x) \ge 0$.
$I(x)$ is interpreted as the probability density for a particle to arrive at 
the screen position $x$. 
According to the usual double slit experiment with plane waves, 
the intensity $I(x)$ should display interference fringes when both slits are 
open, and would be a simple unimodal distribution when only a single slit is 
open. This is what we wanted to test with a more realistic, confined 
geometry. 
The resulting intensities are shown in figs.~\ref{fig:main},~\ref{fig:1slit}(red curves).

Indeed, a very clear (symmetric) interference pattern was found,
with a visibility of the fringes depending on the parameters
of the initial wave-packet. 
This can be heuristically understood as a result
of integrability of the corresponding billiard dynamics.
Namely, the classical ray dynamics inside a $\pi/4$ right triangular
billiard is {\em regular} representing a completely integrable system.
We know that each orbit of an integrable system is characterized by the fact 
that, since the classical motion in $2N$ dimensional phase space is 
confined onto an $N$ invariant torus, at each point in position space, e.g.
at the positions of the slits, only 
a finite number of different momenta (directions) are possible.
Thus the quantum wave-function, in the semiclassical regime,
is expected to be locally a superposition of finitely many 
plane-waves\cite{berry} and the interference pattern on the screen is 
expected to be simply a superposition of fringes using these plane waves.
In our case of an integrable $\pi/4$ right triangular billiard, 
different directions result from specular reflections with the walls.
In contrast to the idealized plane-wave experiment
in infinite domain where interference pattern depends on the
direction of the impact, the fringes here were always symmetric around the
center of the screen. This is a consequence of the presence of the vertical
billiard wall, namely due to collisions with this wall
each impact direction $(v_x,v_y)$ is always accompanied with 
a reflected direction $(-v_x,v_y)$. The pattern on the screen is then a 
{\em symmetric} superposition of the two interference images, one being a 
reflection ($x\to -x$) of the other.
In this way one can also understand that the visibility of the interference
fringes may vary with the direction of the initial packet.

We also remark that the spacing between interference fringes is in agreement
with the usual condition for plane waves that the difference of the distances 
from the two slits to a given point on the screen is an integer multiple 
of $\lambda$.

\begin{figure}
\vspace{-1.5cm}
\centerline{\includegraphics[width=3.3in]{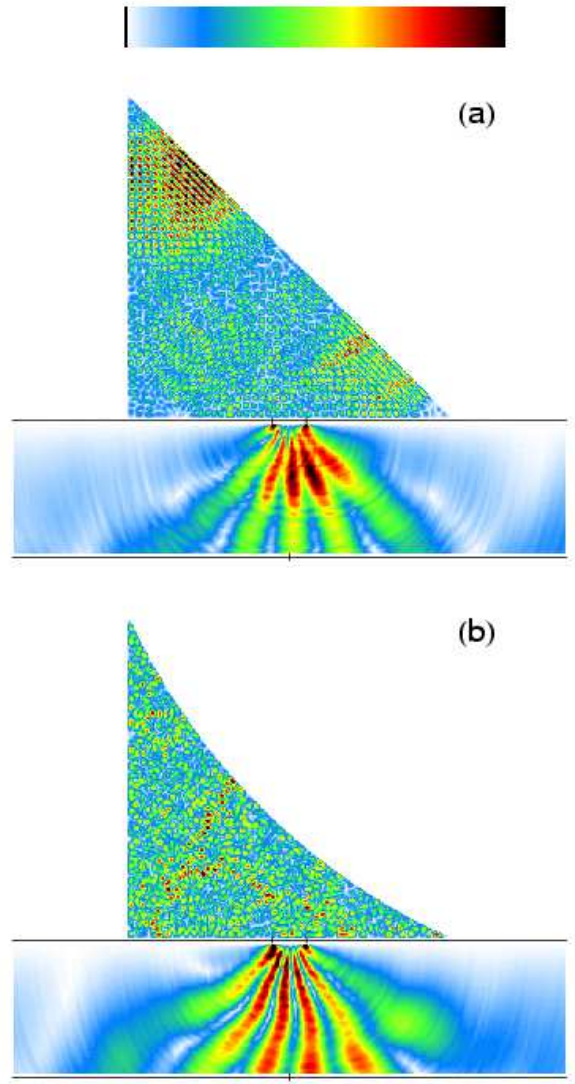}}
\vspace{-1.5cm}
\caption{
Typical snapshots of the wave-function (plotted is the
probability density) for the
two cases: (a) for the regular billiard at $t=0.325$, and (b) for the 
chaotic billiard at $t=0.275$ (both cases correspond to about half the
Heisenberg time). The probability density is normalized separately in
both parts of each plot, namely the probability density, in absolute units, in 
the radiating region is typically less than $1\%$ of the probability density 
in the
billiard domain. The screen, its center, and the positions of the slits
are indicated with thin black lines.
Please note that the color code on the top of the figure is 
proportional to the square root of probability density.
}
\label{fig:wf}
\end{figure}

Now we make a simple modification of our experiment.
We replace the hypotenuse of the triangle by the circular arc of
radius $R=2$ (dashed curve in fig.~\ref{fig:1}). This change has a dramatic consequence for the
classical ray dynamics inside the billiard, namely the latter
becomes fully chaotic. In fact such a dispersive classical billiard is
rigorously known to be a $K$-system \cite{sinai}.
Quite surprisingly, this has also a dramatic effect on the
result of the double slit experiment.
The interference fringes almost completely disappear, and the
intensity can be very accurately reproduced by the sum of intensities
$(I_1(x)+I_2(x))/2$ for the two experiments where only a single
slit is open. This means that the result of such experiment is
the same as would be in terms of classical ray dynamics.
Notice however, that at any given instant of time,
there is a well definite phase relation between the
wave function at both slits. Yet, as time proceeds, this phase
relation changes, and it is lost after averaging over time.
This is nicely illustrated by the snapshots of the wave-functions
in the regular and chaotic case shown in fig.~\ref{fig:wf}.
While in the regular case, the jets of probability emerging from the
slits always point in the same direction and produce a clear
time-integrated fringe structure on the screen, in the chaotic case, the jets
are trembling and moving left and right, thus upon time-integration
they produce no fringes on the screen\cite{commentwigfun}.

\section*{Discussion and conclusions}

The results of this numerical experiment can be understood 
in terms of fast decay of spatial correlations of eigenfunctions
of chaotic systems. In the limit of small slits opening $d \ll \lambda$,
the intensity on the screen, according to simple perturbation expansion
in the small parameter $d/\lambda$, can be written as
\begin{equation}
I(x) = I_1(x) + I_2(x) + C(s)f(x),
\end{equation}
where $f(x)$ is some oscillatory function determining the period of the 
fringes, and $C(s)$ is the spatial correlation function of the normal
derivative of the eigenfunctions $\Psi_n$ of the closed billiard at the
positions $(-s/2,0)$ and $(s/2,0)$ of the slits, written in the 
Cartesian frame
with origin in the middle point between the slits.
In particular, $C(s) = 
\alpha \sum_n |c_n|^2 \partial_y \Psi_n(-s/2,0) \partial_y \Psi_n(s/2,0)$,
where $c_n$ are the expansion coefficients of the initial wave-packet in
the eigenstates $\Psi_n$, and $\alpha$ is a constant such that $C(0) = 1$.
Note that this eigenstate correlation function $C(s)$, which also depends on the
initial state through the expansion coefficients $c_n$, is directly 
proportional to the visibility of the fringes. 
While it is known \cite{berry} that, for classically chaotic
billiards, quantum eigenstates exhibit decaying correlations with
$C(s) = J_0(ks)$ where $J_0$ is a zeroth order Bessel function,
for regular systems  $C(s)$ typically does not decay (but
oscillates) so it produces interference fringes. 
In our case of half-square billiard we find, 
for large $k$, 
$C(s) = e^{-\sigma_k^2 s^2/2}(k_x^2 \cos(k_y s) + k_y^2\cos(k_x s))/k^2$.
The Gaussian prefactor can easily be understood, namely there is no 
interference if the size of the wave-packet is smaller than the slit-distance, or 
equivalently, if uncertainty in momentum $\sigma_k$ is much larger than $1/s$.

Disappearance of interference fringes can be directly related to decoherence. 
If $A$ is a binary observable $A\in\{1,2\}$ which determines 
through which slit the particle went, then $C(s)$ is proportional to the 
non-diagonal matrix element $\bra{1}\rho\ket{2}$
of the density matrix in the eigenbasis of $A$, 
and is thus a direct indicator of decoherence.

In conclusion we have examined the double slit experiment in 
the configuration in which the particle source is confined in 
a two-dimensional billiard region. If the billiard problem is 
classically integrable then interference fringes are observed, as 
in the case of the usual configuration of the gedanken double slit 
experiment with plane waves. However, for a classically chaotic 
billiard, fringes completely disappear and the observed intensity 
on the screen is the sum of the intensities obtained by opening 
one slit at a time.

The result presented here provides therefore, from one hand, a vivid and fundamental illustration of the manifestations of classical chaos in quantum mechanics. On the other hand it shows that, by considering a pure quantum state, in absence of any external decoherence mechanism, internal dynamical chaos can provide the required randomization to ensure quantum to classical transition in the semiclassical region. The effect described in this letter should be observable in a real laboratory experiment.

\section*{Method}
We have implemented an explicit finite difference numerical
method with $\lambda/h \approx 12$ mesh points per de Broglie 
wave-length $\lambda$ where $h$ is the stepsize of the spatial 
discretization.
The stability of the method was enforced by using unitary power-law
expansion of the propagator, namely 
$\Psi(t+\tau) = 
\sum_{j=0}^n \frac{1}{n!}(-\frac{\ii \tau}{\hbar}{\hat H} )^j\Psi(t)
$
where ${\hat H} = -\frac{\hbar^2}{2m}\Delta$ and 
$\Delta$ is a discrete
Laplacian. Using temporal stepsize $\tau = h^2$, the required order $n$ to obtain numerical
convergence within machine precision was typically small, $n < 10$.
The implementation of the finite difference scheme was straightforward for the
triangular geometry, since the boundary conditions conform nicely to the
discretized Cartesian grid. For the case of chaotic billiard,
we used a unique smooth transformation $(x,y)\to (x,f(y))$ which maps the
chaotic billiard geometry to the regular one, and slightly modifies the
calculation of the discrete Laplacian without altering its accuracy 
(due to smoothness of $f(y)$).

\section*{Acknowledgements}

Useful discussion with S. P. Kulik are gratefully acknowledged.
T. P. was financially supported by the grant P1-0044 of the 
Ministry of Science, Education and Sports of Republic of Slovenia.

\end{document}